\documentclass[%
reprint,
superscriptaddress,
amsmath,amssymb,
aps,
prl,
floatfix,
]{revtex4-2}

\usepackage{graphicx}
\graphicspath{{./figures/}}
\usepackage{xcolor}
\usepackage{ulem}
\usepackage{dcolumn}
\usepackage{bm}
\usepackage{hyperref}
\hypersetup{colorlinks, linkcolor = [rgb]{0, 0, 0.5}, citecolor = [rgb]{0,0.0,0.5}, urlcolor = [rgb]{0,0.0,0.5}}
\usepackage[caption=false]{subfig}
\usepackage{siunitx}
\usepackage[capitalise]{cleveref}
\allowdisplaybreaks

\begin{document}

\title{First Extraction of Transverse-Momentum Dependent Helicity Distributions}

\newcommand*{\PKU}{School of Physics, Peking University, Beijing 100871, China}\affiliation{\PKU}
\newcommand*{\CHEP}{Center for High Energy Physics, Peking University, Beijing 100871, China}
\newcommand*{\SDU}{Key Laboratory of Particle Physics and Particle Irradiation (MOE), Institute of Frontier and Interdisciplinary Science, Shandong University, Qingdao, Shandong 266237, China}\affiliation{\SDU}
\newcommand*{\SCNT}{Southern Center for Nuclear-Science Theory (SCNT), Institute of Modern Physics, Chinese Academy of Sciences, Huizhou 516000, China}
\newcommand*{\IMP}{Institute of Modern Physics, Chinese Academy of Sciences, Lanzhou, Gansu Province 730000, China}
\newcommand*{\UCAS}{University of Chinese Academy of Sciences, Beijing 100049, China}
\newcommand*{\ZZU}{School of Physics, Zhengzhou University, Zhengzhou 450001, China}
\newcommand*{\KLQLPIPP}{Key Laboratory of Quark and Lepton Physics (MOE) and Institute of Particle Physics, Central China Normal University, Wuhan 430079, China}

\author{Ke~Yang}
\affiliation{\PKU}
\author{Tianbo~Liu}\email{liutb@sdu.edu.cn}\affiliation{\SDU}\affiliation{\SCNT}
\author{Peng~Sun}\email{pengsun@impcas.ac.cn}\affiliation{\IMP}\affiliation{\UCAS}
\author{Yuxiang~Zhao}\email{yxzhao@impcas.ac.cn}\affiliation{\SCNT}\affiliation{\IMP}\affiliation{\UCAS}\affiliation{\KLQLPIPP}
\author{Bo-Qiang~Ma}\email{mabq@pku.edu.cn}\affiliation{\PKU}\affiliation{\CHEP}\affiliation{\ZZU}

\collaboration{Transverse Nucleon Tomography Collaboration}

\begin{abstract}

We report on the first global analysis of transverse momentum dependent helicity distributions of the proton. The analysis is performed at next-to-leading order with the evolution factor at next-to-next-to-leading-logarithmic accuracy. Nonzero signals are determined for up and down quarks and their $k_T$-integrated polarization are consistent with analyses in collinear factorization, while the distributions of other flavors are loosely constrained by existing data. With increasing transverse momentum, quarks at large $x$ become less polarized while those at small $x$ become more polarized.
 
\end{abstract}

\maketitle


{\it Introduction}---The nucleon spin structure remains one of the most intriguing and unresolved frontiers in hadronic physics~\cite{Aidala:2012mv,Deur:2018roz}. Despite extensive experimental and theoretical studies since the discovery of the proton spin crisis~\cite{EuropeanMuon:1987isl,EuropeanMuon:1989yki}, the precise contributions of quarks and gluons to the proton’s spin are still not fully understood. A critical step in addressing this challenge is the determination of spin-dependent parton distribution functions (PDFs), which is also a central objective of future electron-ion colliders (EICs)~\cite{Accardi:2012qut,AbdulKhalek:2021gbh,Anderle:2021wcy}.

The helicity distribution function $g_1(x)$ represents the difference in density between spin-parallel and spin-antiparallel partons in a longitudinally polarized nucleon, where $x$ denotes the longitudinal momentum fraction carried by the parton. Although first-principle calculations of helicity distributions remain challenging~\cite{Alexandrou:2020uyt,Alexandrou:2021oih,HadStruc:2022yaw,HadStruc:2022nay,Khan:2022vot,Holligan:2024wpv}, they can be extracted from experimental data, such as polarized deep inelastic scattering ~\cite{HERMES:1999uyx,SpinMuon:1997yns,HERMES:2004zsh,COMPASS:2007esq,COMPASS:2009nhs,HERMES:1999yng,HERMES:2010nas,SpinMuonSMC:2004jrx,COMPASS:2012pfa,COMPASS:2012mpe} and $pp$ collisions~\cite{PHENIX:2004aoz,PHENIX:2006nik,PHENIX:2007kqm,PHENIX:2008sgl,PHENIX:2008swq,STAR:2013zyt,PHENIX:2014gbf,PHENIX:2010aru,STAR:2012hth,STAR:2014wox}, based on QCD factorization~\cite{Collins:1989gx,Collins:2011zzd}. Phenomenological analyses of $g_1(x)$ have been conducted by various groups~\cite{deFlorian:2008mr,deFlorian:2009vb,deFlorian:2014yva,Nocera:2014gqa,Blumlein:2002qeu,Blumlein:2010rn,Hirai:2008aj,Leader:2010rb,Sato:2016tuz,Ethier:2017zbq}. Besides the total polarization rate of partons, the ratio between $g_1(x)$ and the unpolarized distribution function $f_1(x)$ reveals the dependence of polarization on parton momentum. Perturbative QCD predicts that the helicity of a parton with a large momentum fraction tends to align with the helicity of the parent nucleon, leading to $g_1(x)/f_1(x)$ approaching $1$ as $x \to 1$, a phenomenon known as helicity retention~\cite{Brodsky:1994kg,Avakian:2007xa}. This behavior is supported by helicity retention implemented analysis~\cite{Leader:2001kh}, statistical picture inspired analysis~\cite{Bourrely:2001du}, and holographic light-front calculations~\cite{Liu:2019vsn}. On the other hand, extrapolation of global analysis to large $x$~\cite{Jimenez-Delgado:2014xza}, Dyson-Schwinger equations~\cite{Roberts:2013mja,Yu:2024qsd}, and quark model calculations~\cite{Ma:1996np,Ma:1997gy,Pasquini:2008ax,Bacchetta:2008af} suggest different values.

The transverse momentum of partons plays an important role in understanding nucleon spin structures. The Wigner rotation effect~\cite{Wigner:1939cj,Melosh:1974cu} states that the polarization of quarks with nonzero transverse momentum $k_T$ will get suppressed when boosting the proton to an infinite momentum frame~\cite{Ma:1991xq,Ma:1992sj}, where the parton language is appropriate~\cite{Feynman:1969ej}. As demonstrated by explicit model calculations~\cite{Ma:1996np,Ma:1996ii}, valence quark helicity distributions qualitatively agree with measurements when taking into account the kinetic effect~\cite{Ma:1996np}, making a crucial step on resolving the proton spin puzzle. However, a complete dynamical boost is impractical due to the nonperturbative nature of QCD at hadronic scales. Valuable insights can be gained from transverse momentum dependent (TMD) helicity distributions $g_{1L}(x,k_T)$, which reveal the polarized density of partons in dependence on their transverse momentum $k_T$ as well as $x$. Therefore, determining TMD helicity distributions provides a crucial perspective on the dynamics of strong interactions, essential for a deeper understanding of nucleon spin structure.

Semi-inclusive deep inelastic scattering (SIDIS) is a main process to extract TMD PDFs, and many analyses have been performed on unpolarized distributions~\cite{Bacchetta:2017gcc,Sun:2014dqm,Scimemi:2019cmh,Moos:2023yfa} and some spin-dependent distributions~\cite{Kang:2015msa,Bacchetta:2020gko,Gamberg:2022kdb,Bury:2021sue,Echevarria:2020hpy,Zeng:2022lbo,Boglione:2024dal,Gamberg:2022kdb,Zeng:2023nnb,Boglione:2024dal,Christova:2020ahe,Bhattacharya:2021twu,Horstmann:2022xkk,Yang:2024bfz}. While TMD physics has become a very active area~\cite{Angeles-Martinez:2015sea,Boussarie:2023izj}, an extraction of TMD helicity distributions remains absent. Recently, the measurements of transverse momentum dependent longitudinal double-spin asymmetry (DSA) in SIDIS were reported by HERMES~\cite{HERMES:2018awh}, COMPASS~\cite{COMPASS:2016klq}, and CLAS~\cite{CLAS:2017yrm}, offering the opportunity toward the determination of TMD helicity distributions.

In this Letter, we perform the first global analysis of TMD helicity distributions of the nucleon. The analysis is carried out within the TMD factorization at next-to-leading order (NLO) and next-to-next-to-leading logarithmic (NNLL) accuracy. Our results reveal that the polarization at large $x$, where the valence components dominate, decreases with increasing transverse momentum, supporting the feature predicted by the Wigner rotation effect. Conversely, in the regime of small $x$, where sea quarks and gluons arising from intricate QCD dynamics dominate, a contrasting $k_T$-dependent behavior is favored. The $k_T$-integrated polarization of $u$- and $d$-quarks is consistent with analyses in collinear factorization, while other flavors are still loosely constrained by existing data, motivating high precision measurements at current and future facilities~\cite{Dudek:2012vr,Accardi:2012qut,AbdulKhalek:2021gbh,Anderle:2021wcy}.

{\it Theoretical framework}---In the SIDIS process,
\begin{align}
    l + P \to l' + P_h + X,
\end{align}
a final-state hadron $h$ is identified in coincidence with the scattered lepton. The labels $l$, $P$, $l'$, and $P_h$ represent the four momenta of the initial-state lepton, the nucleon, the scattered lepton, and the identified hadron respectively, and $X$ stands for the undetected hadronic system. Considering the lepton and the nucleon are both longitudinally polarized, one can express the differential cross section as
\begin{align}
    \frac{d\sigma}{d x d y  d z d P_{hT}^2} 
    &=  
    \frac{4\pi^2\alpha^2}{x y Q^2}\frac{y^2}{2(1-\varepsilon)}\left(1+\frac{\gamma^2}{2x}\right) 
    \notag\\
    &\times 
    \Big[  F_{UU} 
    + \lambda_e S_L \sqrt{1-\varepsilon^2} F_{LL}\Big], 
    \label{eq:CSLT}
\end{align}
where
\begin{align}
    Q^2 & = -q^2 = -(l-l')^2, \\
    x &= \frac{Q^2}{2P\cdot q},
    \quad
    y = \frac{P\cdot q}{P\cdot l},
    \quad
    z = \frac{P\cdot P_h}{P\cdot q}, \\
    \gamma &= \frac{2xM}{Q} = \frac{MQ}{P\cdot q},
\end{align}
$M$ is the nucleon mass, $\alpha$ is the electromagnetic fine structure constant, $\lambda_e$ represents the helicity of the lepton beam, and $S_L$ represents the longitudinal polarization of the nucleon. The ratio of longitudinal and transverse photon flux is given by
\begin{align}
    \varepsilon = \frac{1-y-\frac{1}{4}\gamma^2 y^2}{1-y+\frac{1}{2}y^2 + \frac{1}{4}\gamma^2 y^2}.
\end{align}
The transverse momentum $P_{hT}$ of the hadron is defined in the virtual photon-nucleon frame, the same as the Trento conventions~\cite{Bacchetta:2004jz}.

In this process, the large momentum transfer mediated by the virtual photon provides a hard scale $Q \gg \Lambda_{\rm QCD}$, serving as a probe to the partons in the nucleon. A second scale is characterized by the hadron transverse momentum $P_{hT}$. When $P_{hT} \ll Q$, the cross section is sensitive to the intrinsic transverse motion of partons, and one can apply the TMD factorization~\cite{Collins:1981uk,Ji:2004wu,Ji:2004xq,Aybat:2011zv}. Then the structure functions, such as $F_{UU}$ and $F_{LL}$ in~\eqref{eq:CSLT}, are expressed as convolutions of TMD PDFs and TMD fragmentation functions (FFs).

In perturbative expansion in power of the strong coupling constant $\alpha_s$, one will find logarithmic enhanced terms at each fixed order. For a reliable prediction, one needs to resum the large logarithms of all orders into an evolution factor. It is convenient to implement the TMD evolution in the $b$ space, the Fourier transform of the $k_T$ space, and the scale dependence is determined by the equations~\cite{Aybat:2011zv,Scimemi:2018xaf}
\begin{align}
    \mu \frac{d {\cal F}(x,b;\mu,\zeta)}{d\mu} & = \gamma_F(\mu,\zeta) {\cal F}(x,b;\mu,\zeta), \\
    \zeta \frac{d {\cal F}(x,b;\mu,\zeta)}{d\zeta} &= -{\cal D}(\mu,b) {\cal F}(x,b;\mu,\zeta),
\end{align}
where $\mu$ is the renormalization scale and $\zeta$ serves as a cutoff scale to regularize the light cone singularity. $\gamma_F$ is the TMD anomalous dimension, $\cal D$ is the rapidity anomalous dimension, and $\cal F$ represents a TMD PDF or FF. By solving the evolution equations, one can formally relate the TMD functions at different scales as
\begin{align}
    {\cal F}(x,b;\mu_f,\zeta_f) = R[b;(\mu_i,\zeta_i)\to(\mu_f,\zeta_f)] {\cal F}(x,b;\mu_i,\zeta_i),
\end{align}
where the evolution factor is a path integral from $(\mu_i,\zeta_i)$ to $(\mu_f,\zeta_f)$.
According to the integrability condition~\cite{Collins:1981va}, the evolution factor is in principle path independent. However, it differs from path to path when truncating the perturbative series. In this analysis, we follow the $\zeta$ prescription~\cite{Scimemi:2019cmh,Moos:2023yfa}, which was also adopted in recent studies of Sivers distributions~\cite{Bury:2021sue,Zeng:2022lbo}, the transversity distributions~\cite{Zeng:2023nnb}, and the worm-gear distributions~\cite{Horstmann:2022xkk,Yang:2024bfz}. The initial scales are set at the saddle point in the $(\mu,\zeta)$ plane defined by
\begin{align}
    {\cal D}(\mu_i,b) = 0,
    \quad
    \gamma_F(\mu_i,\zeta_i) = 0,
    \label{eq:saddle}
\end{align}
where the $\cal F$ is referred to as the optimal TMD function. The corresponding TMD function at $(\mu_f,\zeta_f)$ is obtained by evolving the optimal function $\cal F$ along the equipotential line $\zeta_\mu(\mu,b)$ determined by
\begin{align}
    \frac{d \ln \zeta_\mu(\mu,b)}{d\ln \mu^2} 
    =
    \frac{\gamma_F(\mu,\zeta_\mu(\mu,b))}{2{\cal D}(\mu,b)},
    \label{eq:equipotential}
\end{align}
and then along the straight line to $\zeta_f$. As a common choice, we set $\mu_f^2 = \zeta_f = Q^2$, and the evolution factor is
\begin{align}
    R[b;(\mu_i,\zeta_i)\to(Q,Q^2)] = \Bigg[\frac{Q^2}{\zeta_\mu(Q,b)}\Bigg]^{-{\cal D}(Q,b)}.
\end{align}

According to the TMD factorization, the TMD helicity distributions can be extracted from the measurements of longitudinal DSA,
\begin{align}
    A_{LL} = \frac{\sqrt{1-\varepsilon^2} F_{LL}(x,z,P_{hT}^2,Q^2)}{F_{UU}(x,z,P_{hT}^2,Q^2)},
\end{align}
at low transverse momentum. The structure functions are expressed in terms of TMD PDFs and FFs as
\begin{align}
    F_{UU} &= |C_V(Q^2,\mu)|^2 x \sum_q e_q^2 \int_0^\infty \frac{db}{2\pi} b J_0\Big(\frac{b P_{hT}}{z}\Big) \notag\\
    &\times
    f_{1}^q(x,b;Q,Q^2) D_{1}^{q\to h}(z,b;Q,Q^2), \\
    F_{LL} &= |C_V(Q^2,\mu)|^2 x \sum_q e_q^2 \int_0^\infty \frac{db}{2\pi} b J_0\Big(\frac{b P_{hT}}{z}\Big) \notag\\ 
    &\times
    g_{1L}^q(x,b;Q,Q^2) D_{1}^{q\to h}(z,b;Q,Q^2),
\end{align}
where $C_V$ is a hard factor of partonic scatterings that can be calculated perturbatively, $e_q$ is the charge of the corresponding parton flavor, and $J_0$ is the zeroth Bessel function arising from the transverse Fourier transform. $f_1$, $g_{1L}$, and $D_1$ are unpolarized TMD PDF, helicity TMD PDF, and unpolarized TMD FF, respectively, in the $b$ space, with superscript $q$ labeling the parton flavor.

{\it Parametrization and analysis}---With the formalism above, we perform a global analysis of the world SIDIS DSA data reported by HERMES~\cite{HERMES:2018awh} and CLAS~\cite{CLAS:2017yrm}. Since kinematics information of the COMPASS data is incomplete~\cite{COMPASS:2016klq}, they are not included in this analysis.

We parametrize the optimal TMD helicity distributions, {\it i.e.}, at the saddle point given by Eq.~\eqref{eq:saddle}, as
\begin{align}
    g_{1L}(x,b) &= \sum_{f'} \int_x^1 \frac{d\xi}{\xi} \Delta C_{f\leftarrow f'}\big(\xi,b,\mu_{\rm OPE}\big)  \notag \\
    &\quad\times
    g_{1L}^{f'}\Big(\frac{x}{\xi}\Big)g_{\rm NP}(x,b),
\end{align}
where
\begin{align}
\label{eq:g1L(x)}
    g_{1L}^f(x) &= N_f\frac{(1-x)^{\alpha_f} x^{\beta_f} (1+\epsilon_f x)}{n(\alpha_f,\beta_f,\epsilon_f)}
    g_1^f(x,\mu_{\rm OPE}),\\
    g_{\rm NP}(x,b) &=\exp \left[ -\frac{\lambda_1 (1-x) + \lambda_2 x + \lambda_5 x (1-x)}{\sqrt{1+\lambda_3 x^{\lambda_4}b^2}}b^2 \right],
\end{align}
with $g_1(x,\mu_{\rm OPE})$ taken from the NNPDFpol1.1~\cite{Nocera:2014gqa} at $\mu_{\rm OPE}$, and $N_f$, $\alpha_f$, $\beta_f$, $\epsilon_f$, and $\lambda_i$ being parameters to fit. The factor $n(\alpha,\beta,\epsilon) = (\alpha + \beta + 6 + 2\epsilon + \beta\epsilon)B(\alpha+4,\beta+2)/(\alpha+\beta+6)$ is introduced to reduce the correlation among parameters. 
The $x$ dependent factor in Eq. (17) allows a variation from the collinear distribution. Such an $x$-shape modification is removed if setting $\alpha_{f} = \beta_{f} = \epsilon_{f} = 0$.
The coefficients $\Delta C_{f\leftarrow f'}$ are obtained from the small-$b$ operator product expansion (OPE) and explicit expressions up to NLO can be found in Ref.~\cite{Gutierrez-Reyes:2017glx}, and $\mu_{\rm OPE}$ is chosen as $2e^{-\gamma_E}/b + 2\,\rm GeV$. 
For unpolarized TMD PDFs $f_1$ and TMD FFs $D_1$, we adopt the simultaneous fit SV19 ~\cite{Scimemi:2019cmh} by global analysis of Drell-Yan and SIDIS data using NNPDF31 collinear PDF set~\cite{NNPDF:2017mvq} and collinear FF set in Refs.~\cite{deFlorian:2014xna,deFlorian:2017lwf,deFlorian:2007ekg}. 
They were also extracted within the $\zeta$-prescription and have been utilized in recent TMD analyses~\cite{Bury:2021sue,Zeng:2022lbo,Zeng:2023nnb,Horstmann:2022xkk,Yang:2024bfz}.

Since the TMD factorization is valid at low transverse momentum, we impose the cuts $P_{hT}/(zQ) < 0.5$ and $Q > 1\,\rm GeV$, resulting in 253 data points included in the fit, as listed in Table~\ref{tab:data}. According to the amount and precision of existing world data, which cannot efficiently constrain a huge number of parameters, we set $\alpha_f = \beta_f = \epsilon_f = 0$ for $\bar{u}$, $\bar{d}$, $s$, $\bar{s}$, and $g$, and $N_s = N_{\bar s}$ in this analysis. Then there are in total 17 free parameters.

\begin{table}
\caption{SIDIS DSA datasets in the analysis. The numbers in parentheses are data points before the $P_{hT}$ cut. The last column provides the $\chi^2$ per data points for each dataset.}
\label{tab:data}
\begin{ruledtabular}
\begin{tabular}{llll}
Experiment  &  Process       & Data points   & $\chi^2/N$      \\ \hline 
HERMES\cite{HERMES:2018awh}   & $e^{\pm}p\rightarrow e^{\pm} hX$ & 84 (160) & 0.72 \\
HERMES\cite{HERMES:2018awh}   & $e^{\pm}d\rightarrow e^{\pm} hX$ & 160 (317) & 0.71 \\
CLAS\cite{CLAS:2017yrm}   & $e^{-}p\rightarrow e^{-} \pi^{0}X$ & 9 (21) & 1.43 \\
\hline
Total           &     &     253 (498) &   $0.74$
\end{tabular}
\end{ruledtabular}
\end{table}

The discrepancy between theoretical predictions and experimental measurements is quantified by
\begin{align}
    \chi^2 = \sum_{\rm sets} \sum_{i,j} (t_i - a_i) V^{-1}_{ij} (t_j - a_j),
\end{align}
where the first summation runs over all datasets and the second summation runs over data points in a dataset. Within each dataset, $t_i$ and $a_i$ are respectively the calculated and the measured values of the $i$th point, and $V_{ij}$ is the covariance matrix, which contains data uncertainties and their correlations.

\begin{figure*}[htp]
    \centering
    \includegraphics[width=0.3\textwidth]{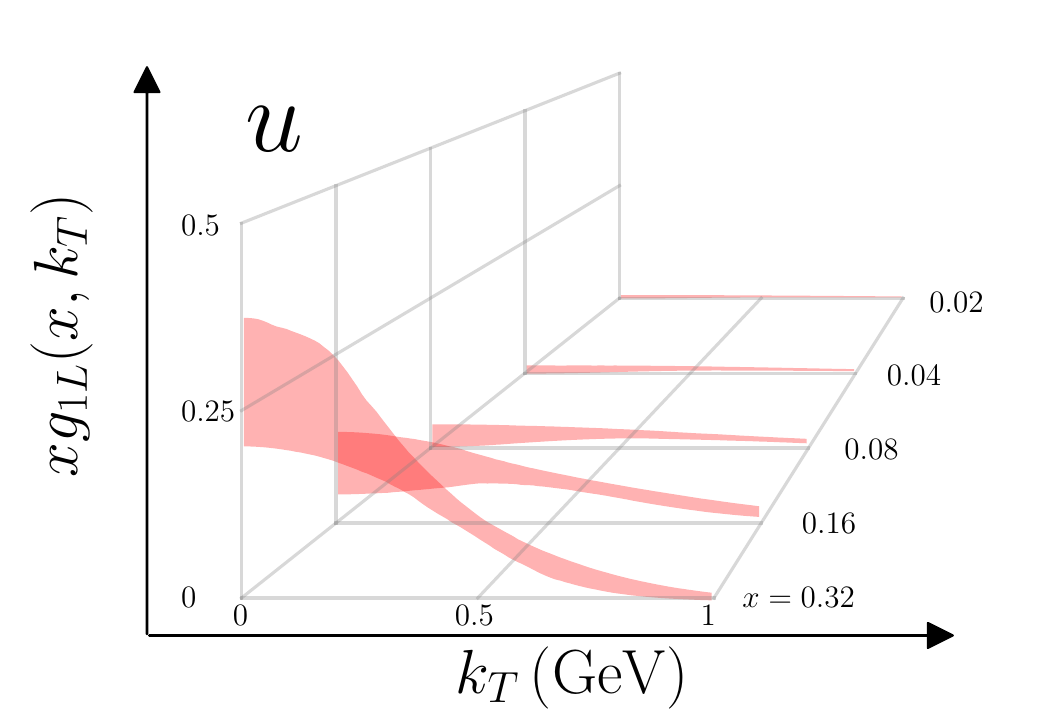}
    \includegraphics[width=0.3\textwidth]{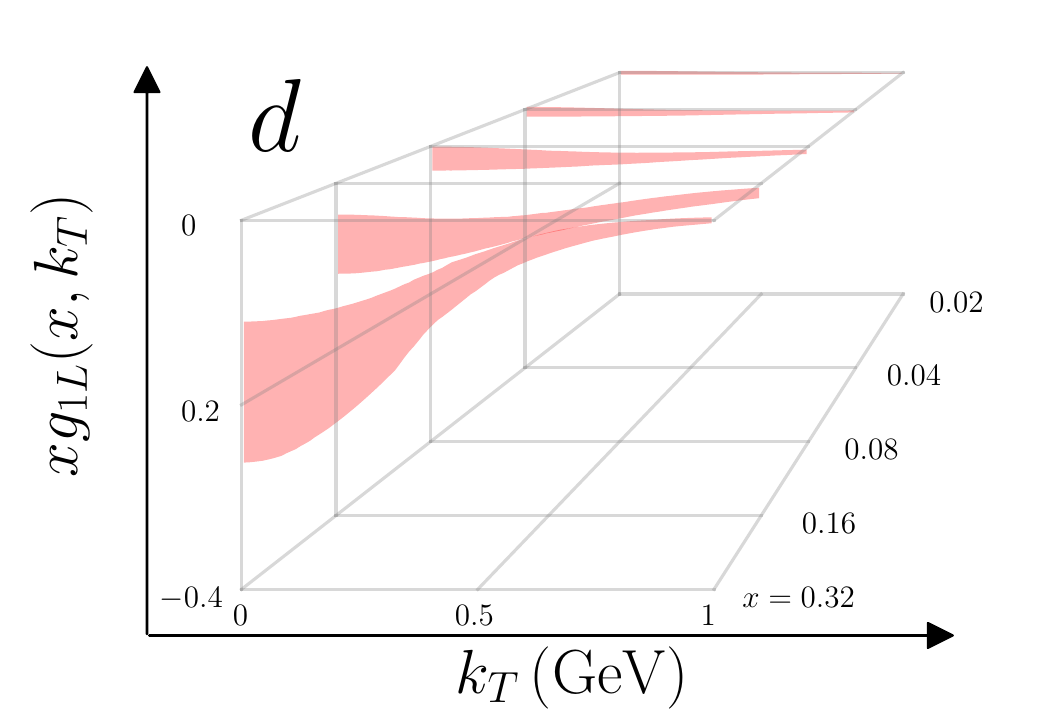}
    \includegraphics[width=0.3\textwidth]{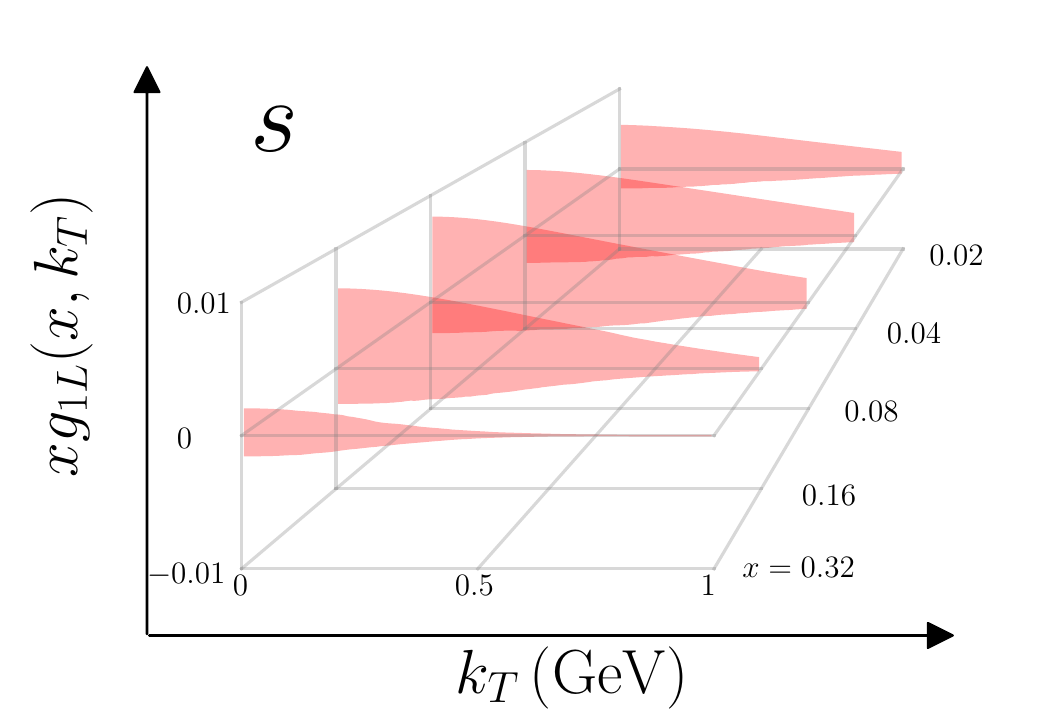} \\
    \includegraphics[width=0.3\textwidth]{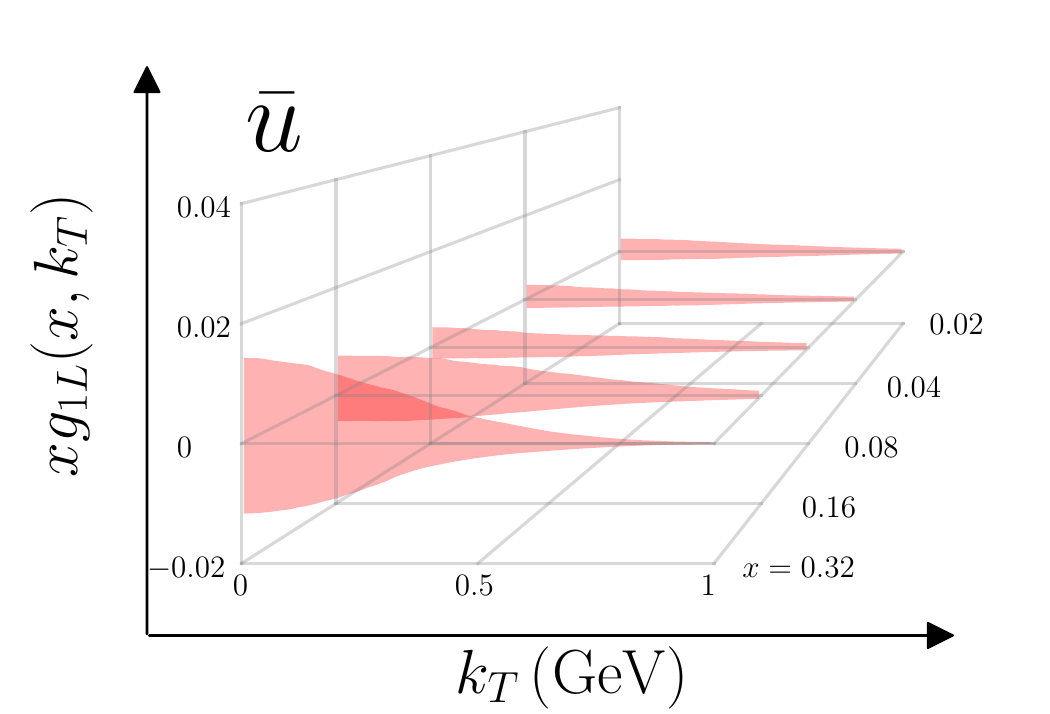}
    \includegraphics[width=0.3\textwidth]{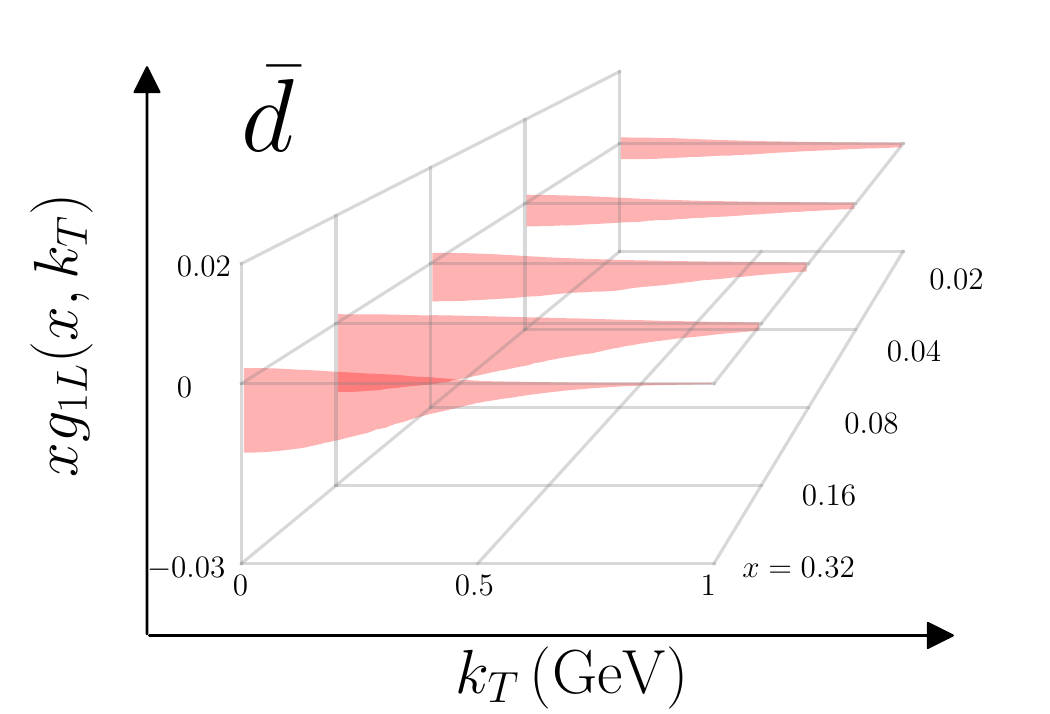}
    \includegraphics[width=0.3\textwidth]{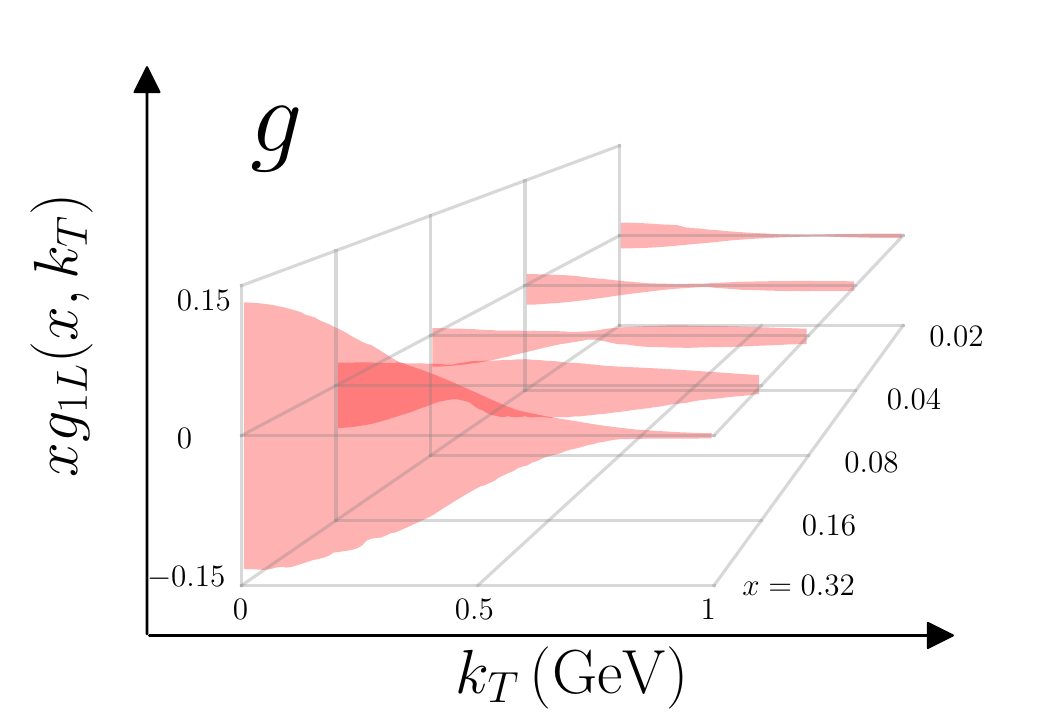}
    \caption{Results of TMD helicity distributions at $Q=2\,\rm GeV$. The bands represent 68\% CL from the fits of 1000 replicas.}
    \label{fig:xg1L_3d}
\end{figure*}

To estimate the uncertainties of extracted distributions, we produce 1000 replicas taking into account the uncertainties and correlations of data points. The fitted values, uncertainties, and the correlation matrix of the parameters are provided in the Supplemental Material~\cite{sup}. We note that the $\chi^2/N$ value for the CLAS data set is a bit large. As the CLAS measurement is at a lower energy than that of the HERMES measurement, power corrections may have sizable contributions. To further examine this effect, high precision data in a wide kinematic coverage are desired.
The results of extracted TMD helicity distributions $g_{1L}(x,k_T)$ are shown in Fig.~\ref{fig:xg1L_3d}, in which the bands represent 68\% CL around the averaged results from all replicas. As one can observe from the figure, positive $u$-quark TMD helicity distribution and negative $d$-quark TMD helicity distribution are determined with clear nonzero signals, although the distributions of sea quarks and gluons are loosely constrained because of the limited $P_{hT}$-dependent $A_{LL}$ data. Future polarized SIDIS experiments at EICs~\cite{Accardi:2012qut,AbdulKhalek:2021gbh,Anderle:2021wcy} are expected to provide high precision data to constrain the distributions of sea quarks and gluons.

\begin{figure}[htp]
    \centering
    \includegraphics[width=0.85\columnwidth]{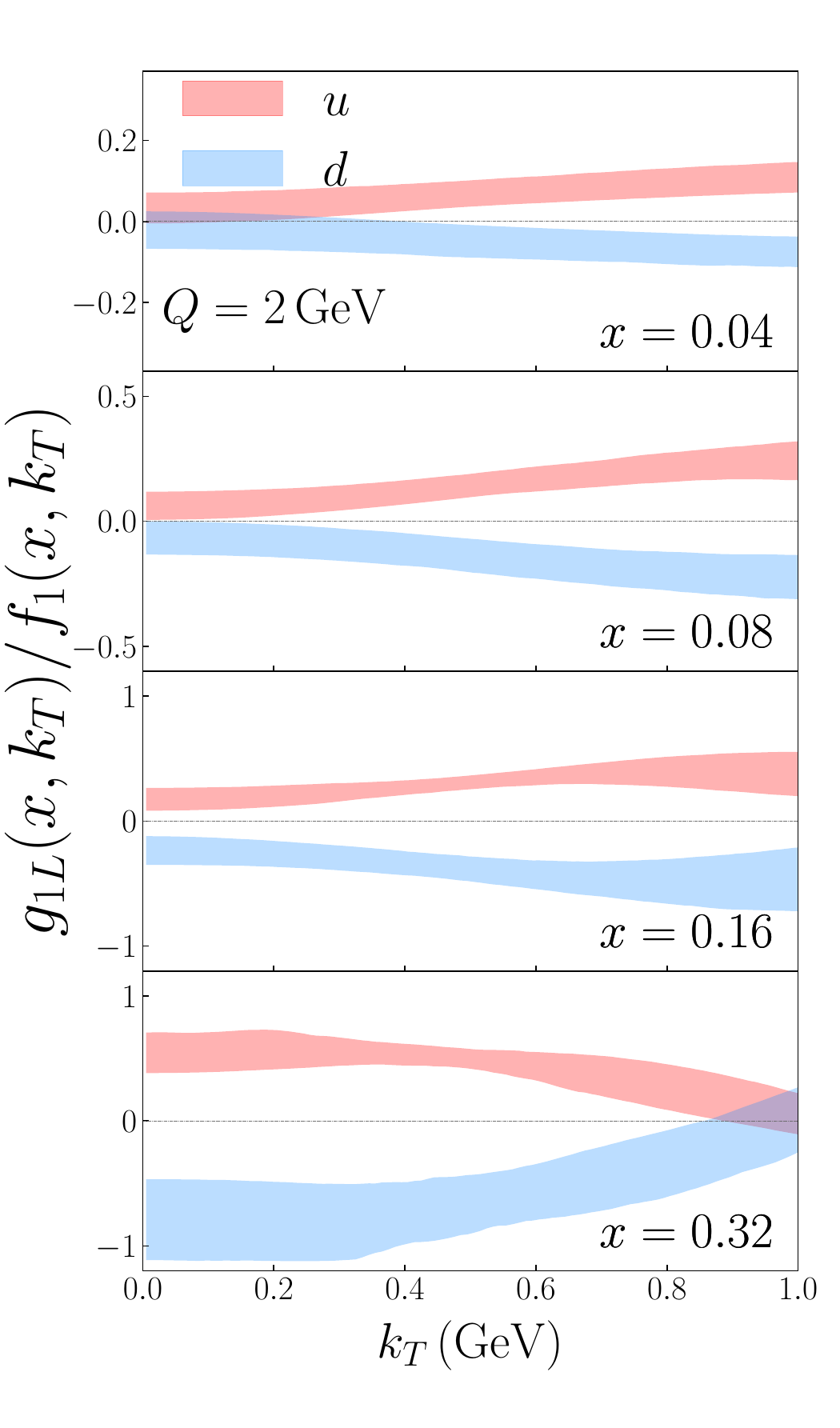}
    \caption{Transverse momentum dependence of the polarization of $u$ quark (red) and $d$ quark (blue). The bands represent 68\% CL.}
    \label{fig:polarization_x_kt}
\end{figure}

Apart from the absolute TMD helicity distributions, it is interesting to examine the ratio $g_{1L}(x,k_T)/f_1(x,k_T)$, which reveals the polarization of partons induced by the polarization of the parent proton. As shown in Fig.~\ref{fig:polarization_x_kt}, at large $x$, where the valence component dominates, the polarization of both $u$ and $d$ quarks decreases with increasing $k_T$. This feature is qualitatively consistent with the kinetic Wigner rotation effect~\cite{Wigner:1939cj,Melosh:1974cu,Ma:1991xq,Ma:1992sj}. However, a contrasting behavior that the polarization increases with $k_T$ is observed at relatively low $x$ values. In this region, the valence component is no longer adequate and parton distributions are highly driven by complex QCD dynamics. Therefore, valuable insights on nucleon spin structures and strong interaction dynamics can be obtained from TMD helicity distributions.

\begin{figure}[htp]
    \centering
    \includegraphics[width=0.85\columnwidth]{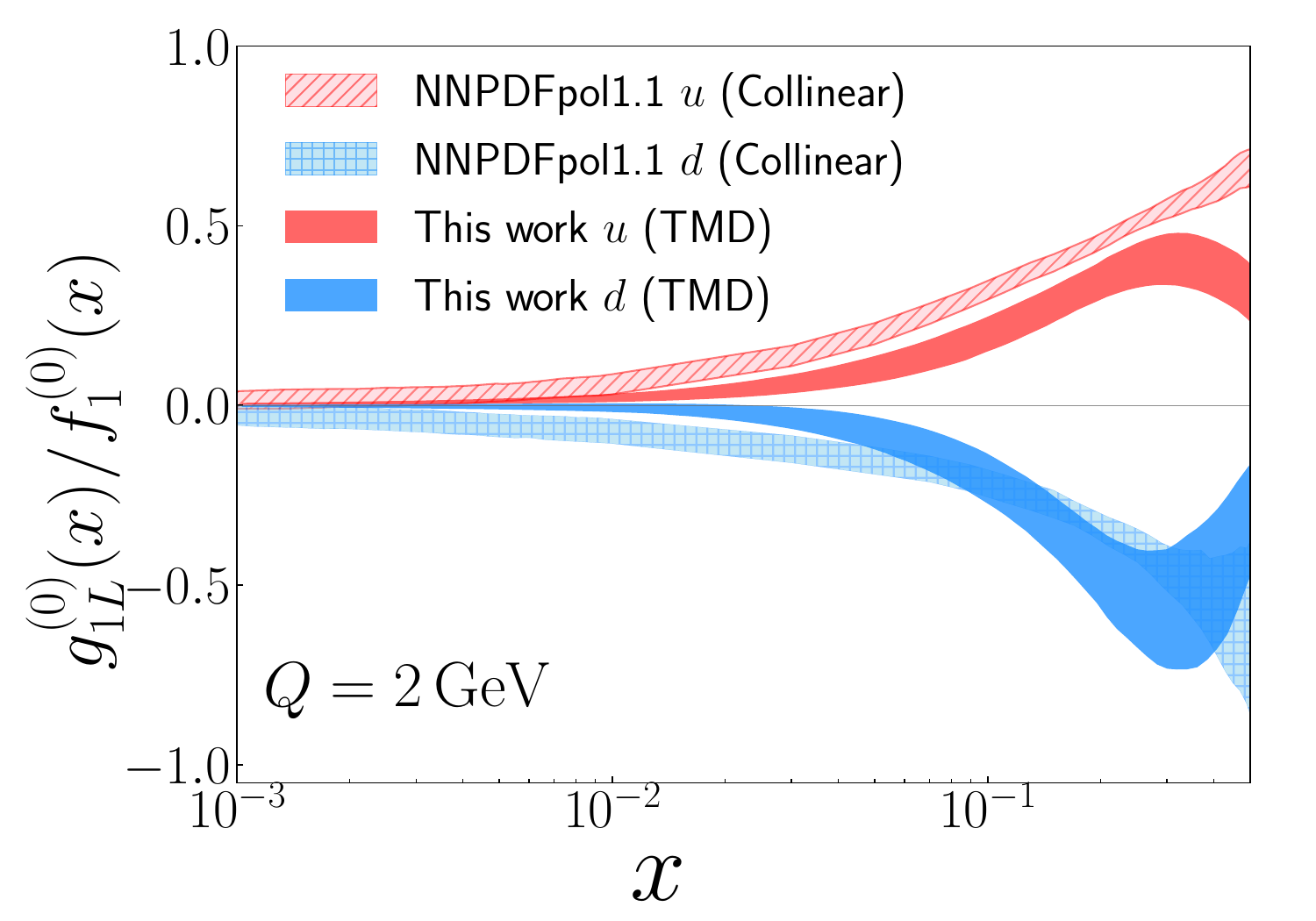}
    \caption{Transverse momentum integrated polarization distributions of $u$ quark (red) and $d$ quark (blue) and the comparison with those from collinear distributions~\cite{Nocera:2014gqa,NNPDF:2017mvq}.}
    \label{fig:polarization_x}
\end{figure}

In addition, we also calculate the $k_T$-integrated distributions,
\begin{align}
    g_{1L}^{(0)}(x) = \int d^2 k_T\, g_{1L}(x,k_T),
\end{align}
which is also referred to as the zeroth transverse momentum moment. Here we differentiate the notation from the collinear helicity distribution $g_1(x)$, since the bare level identity between $k_T$ integrated TMD distribution and collinear distribution does not hold at the renormalized level~\cite{Echevarria:2011epo,Collins:2011zzd}. Despite this fact, a numerical examination of unpolarized TMD and collinear distributions suggested that approximate agreement might be achieved if applying a cut of the $k_T$-integral up to some value $k_T^{\rm max} \sim Q$~\cite{Ebert:2022cku}. Hence we apply the cut to evaluate the $k_T$ integral and vary $k_T^{\rm max}$ from $1$ to $2\,\rm GeV$ as part of the uncertainties. The $k_T$-integrated polarization distributions are shown in Fig.~\ref{fig:polarization_x}, in comparison with those from collinear analysis~\cite{Nocera:2014gqa,NNPDF:2017mvq}. Within the data covered region, up to $x \sim 0.3$, the TMD and collinear results roughly agree with each other, while there are deviations when extrapolating to a higher $x$ region. Polarized SIDIS experiments at Jefferson Lab can make measurements at larger $x$ values~\cite{Dudek:2012vr}, which will improve the determination in the extrapolated region.

{\it Summary and outlook}---We report the first global analysis of TMD helicity distributions. The analysis is performed within the TMD factorization at NLO and NNLL accuracy by fitting the longitudinal DSA measurements in the SIDIS process. The results show nonzero signals of $u$ quark and $d$ quark TMD helicity distributions, and their $k_T$-integrated polarization distributions 
are compatible with collinear PDF extractions across the range of $x$ values covered by the data.

In addition to the $x$-dependence, the TMD helicity distributions reveal the $k_T$-dependence of parton polarization induced by the nucleon polarization, providing information that is not captured in collinear distributions.
Our results indicate that both $u$-quark and $d$-quark polarization decreases with $k_T$ in the valence component dominant region. This behavior qualitatively matches the feature predicted by the Wigner rotation effect. On the other hand, in the relative low-$x$ region, where QCD dynamics plays an essential role, increasing polarization in dependence on $k_T$ is observed. In either case, the $k_T$-dependent polarization is highly nontrivial, and will deepen our understanding of nucleon spin structures and shed light on the dynamics of strong interaction. So more scrutiny is desired for TMD helicity distributions.

{\it Acknowledgments}---The authors acknowledge the computing resource at the Southern Nuclear Science Computing Center and the High-performance Computing Platform of Peking University.
T.L. is supported by the National Natural Science Foundation of China under grants No.~12175117 and No.~12321005, and
Shandong Province Natural Science Foundation Grant
No. ZFJH202303.
P.S. is supported by the Natural Science Foundation of China under grants No.~11975127 and No.~12061131006.
Y.Z. is supported by the Strategic Priority Research Program of the Chinese Academy of Sciences under grant No.~XDB34000000, the Guangdong Provincial Key Laboratory of Nuclear Science under grant No. 2019B121203010, and the Guangdong Major Project of Basic and Applied Basic Research No.~2020B0301030008.
B.-Q.M. is supported by the National Natural Science Foundation of China under grants No.~12075003 and No.~12335006.

{\it Note added}---There is another analysis~\cite{Bacchetta:2024yzl} appeared during the reviewing process of this work.


\begin{thebibliography}{}

\bibitem{Aidala:2012mv}
C.~A.~Aidala, S.~D.~Bass, D.~Hasch, and G.~K.~Mallot,
The spin structure of the nucleon,
\href{https://doi.org/10.1103/RevModPhys.85.655}{Rev. Mod. Phys. \textbf{85}, 655 (2013)}.

\bibitem{Deur:2018roz}
A.~Deur, S.~J.~Brodsky, and G.~F.~de T\'eramond,
The spin structure of the nucleon,
\href{https://doi.org/10.1088/1361-6633/ab0b8f}{Rep. Prog. Phys. \textbf{82}, 076201 (2019)}.

\bibitem{EuropeanMuon:1987isl}
J.~Ashman \textit{et al.} (European Muon Collaboration),
A measurement of the spin asymmetry and determination of the structure function $\textsl{g}_1$ in deep inelastic muon-proton scattering,
\href{https://doi.org/10.1016/0370-2693(88)91523-7}{Phys. Lett. B \textbf{206}, 364 (1988)}.

\bibitem{EuropeanMuon:1989yki}
J.~Ashman \textit{et al.} (European Muon Collaboration),
An investigation of the spin structure of the proton in deep inelastic scattering of polarised muons on polarised protons,
\href{https://doi.org/10.1016/0550-3213(89)90089-8}{Nucl. Phys. B \textbf{328}, 1 (1989)}.

\bibitem{Accardi:2012qut}
A.~Accardi \textit{et al.},
Electron-Ion Collider: The next QCD frontier: Understanding the glue that binds us all,
\href{https://doi.org/10.1140/epja/i2016-16268-9}{Eur. Phys. J. A \textbf{52}, no.9, 268 (2016)}.

\bibitem{AbdulKhalek:2021gbh}
R.~Abdul Khalek \textit{et al.},
Science Requirements and Detector Concepts for the Electron-Ion Collider: EIC Yellow Report,
\href{https://doi.org/10.1016/j.nuclphysa.2022.122447}{Nucl. Phys. A \textbf{1026}, 122447 (2022)}.

\bibitem{Anderle:2021wcy}
D.~P.~Anderle \textit{et al.},
Electron-ion collider in China,
\href{https://doi.org/10.1007/s11467-021-1062-0}{Front. Phys. (Beijing) \textbf{16}, no.6, 64701 (2021)}.

\bibitem{Alexandrou:2020uyt}
C.~Alexandrou, M.~Constantinou, K.~Hadjiyiannakou, K.~Jansen, and F.~Manigrasso,
Flavor decomposition for the proton helicity parton distribution functions,
\href{https://doi.org/10.1103/PhysRevLett.126.102003}{Phys. Rev. Lett. \textbf{126}, no.10, 102003 (2021)}.

\bibitem{Alexandrou:2021oih}
C.~Alexandrou, M.~Constantinou, K.~Hadjiyiannakou, K.~Jansen, and F.~Manigrasso,
Flavor decomposition of the nucleon unpolarized, helicity, and transversity parton distribution functions from lattice QCD simulations,
\href{https://doi.org/10.1103/PhysRevD.104.054503}{Phys. Rev. D \textbf{104}, no.5, 054503 (2021)}.

\bibitem{HadStruc:2022yaw}
C.~Egerer \textit{et al.} (HadStruc Collaboration),
Toward the determination of the gluon helicity distribution in the nucleon from lattice quantum chromodynamics,
\href{https://doi.org/10.1103/PhysRevD.106.094511}{Phys. Rev. D \textbf{106}, no.9, 094511 (2022)}.

\bibitem{HadStruc:2022nay}
R.~G.~Edwards \textit{et al.} (HadStruc Collaboration),
Non-singlet quark helicity PDFs of the nucleon from pseudo-distributions,
\href{https://doi.org/10.1007/JHEP03(2023)086}{J. High Energy Phys. 03 (2023) 086}.

\bibitem{Khan:2022vot}
T.~Khan, T.~Liu, and R.~S.~Sufian,
Gluon helicity in the nucleon from lattice QCD and machine learning,
\href{https://doi.org/10.1103/PhysRevD.108.074502}{Phys. Rev. D \textbf{108}, no.7, 074502 (2023)}.

\bibitem{Holligan:2024wpv}
J.~Holligan and H.~W.~Lin,
Nucleon helicity parton distribution function in the continuum limit with self-renormalization,
\href{https://doi.org/10.1016/j.physletb.2024.138731}{Phys. Lett. B \textbf{854}, 138731 (2024)}.

\bibitem{HERMES:1999uyx}
K.~Ackerstaff \textit{et al.} (HERMES Collaboration),
Flavor decomposition of the polarized quark distributions in the nucleon from inclusive and semiinclusive deep inelastic scattering,
\href{https://doi.org/10.1016/S0370-2693(99)00964-8}{Phys. Lett. B \textbf{464}, 123 (1999)}.

\bibitem{SpinMuon:1997yns}
B.~Adeva \textit{et al.} (Spin Muon Collaboration),
Polarized quark distributions in the nucleon from semiinclusive spin asymmetries,
\href{https://doi.org/10.1016/S0370-2693(97)01546-3}{Phys. Lett. B \textbf{420}, 180 (1998)}.

\bibitem{HERMES:2004zsh}
A.~Airapetian \textit{et al.} (HERMES Collaboration),
Quark helicity distributions in the nucleon for up, down, and strange quarks from semi-inclusive deep-inelastic scattering,
\href{https://doi.org/10.1103/PhysRevD.71.012003}{Phys. Rev. D \textbf{71}, 012003 (2005)}.

\bibitem{COMPASS:2007esq}
M.~Alekseev \textit{et al.} (COMPASS Collaboration),
The polarised valence quark distribution from semi-inclusive DIS,
\href{https://doi.org/10.1016/j.physletb.2007.12.056}{Phys. Lett. B \textbf{660}, 458 (2008)}.

\bibitem{COMPASS:2009nhs}
M.~Alekseev \textit{et al.} (COMPASS Collaboration),
Measurement of the longitudinal spin transfer to $\Lambda$ and $\bar{\Lambda}$ hyperons in polarised muon DIS,
\href{https://doi.org/10.1140/epjc/s10052-009-1143-7}{Eur. Phys. J. C \textbf{64}, 171 (2009)}.

\bibitem{HERMES:1999yng}
A.~Airapetian \textit{et al.} (HERMES Collaboration),
Measurement of the Spin Asymmetry in the Photoproduction of Pairs of High-$p_T$ Hadrons at HERMES,
\href{https://doi.org/10.1103/PhysRevLett.84.2584}{Phys. Rev. Lett. \textbf{84}, 2584 (2000)}.

\bibitem{HERMES:2010nas}
A.~Airapetian \textit{et al.} (HERMES Collaboration),
Leading-order determination of the gluon polarization from high-$p_T$ hadron electroproduction,
\href{https://doi.org/10.1007/JHEP08(2010)130}{J. High Energy Phys. 08 (2010) 130}.

\bibitem{SpinMuonSMC:2004jrx}
B.~Adeva \textit{et al.} (Spin Muon Collaboration),
Spin asymmetries for events with high $p_T$ hadrons in DIS and an evaluation of the gluon polarization,
\href{https://doi.org/10.1103/PhysRevD.70.012002}{Phys. Rev. D \textbf{70}, 012002 (2004)}.

\bibitem{COMPASS:2012pfa}
C.~Adolph \textit{et al.} (COMPASS Collaboration),
Leading order determination of the gluon polarisation from DIS events with high-$p_T$ hadron pairs,
\href{https://doi.org/10.1016/j.physletb.2012.11.056}{Phys. Lett. B \textbf{718}, 922 (2013)}.

\bibitem{COMPASS:2012mpe}
C.~Adolph \textit{et al.} (COMPASS Collaboration),
Leading and next-to-leading order gluon polarization in the nucleon and longitudinal double spin asymmetries from open charm muoproduction,
\href{https://doi.org/10.1103/PhysRevD.87.052018}{Phys. Rev. D \textbf{87}, no.5, 052018 (2013)}.

\bibitem{PHENIX:2004aoz}
S.~S.~Adler \textit{et al.} (PHENIX Collaboration),
Double Helicity Asymmetry in Inclusive Midrapidity $\pi^0$ Production for Polarized $p + p$ collisions at $\sqrt{s} = 200~\rm GeV$,
\href{https://doi.org/10.1103/PhysRevLett.93.202002}{Phys. Rev. Lett. \textbf{93}, 202002 (2004)}.

\bibitem{PHENIX:2006nik}
S.~S.~Adler \textit{et al.} (PHENIX Collaboration),
Improved measurement of double helicity asymmetry in inclulsive midrapidity $\pi^0$~production for polarized $p+p$~collisions at $\sqrt{s} = 200$ GeV,
\href{https://doi.org/10.1103/PhysRevD.73.091102}{Phys. Rev. D \textbf{73}, 091102 (2006)}.

\bibitem{PHENIX:2007kqm}
A.~Adare \textit{et al.} (PHENIX Collaboration),
Inclusive cross-section and double helicity asymmetry for $\pi^0$ production in $p + p$ collisions at $\sqrt{s} =$ 200 GeV: Implications for the polarized gluon distribution in the proton,
\href{https://doi.org/10.1103/PhysRevD.76.051106}{Phys. Rev. D \textbf{76}, 051106 (2007)}.

\bibitem{PHENIX:2008sgl}
A.~Adare \textit{et al.} (PHENIX Collaboration),
Inclusive cross section and double helicity asymmetry for $\pi^0$ production in $p + p$ collisions at $\sqrt{s}=62.4$ GeV,
\href{https://doi.org/10.1103/PhysRevD.79.012003}{Phys. Rev. D \textbf{79}, 012003 (2009)}.

\bibitem{PHENIX:2008swq}
A.~Adare \textit{et al.} (PHENIX Collaboration),
Gluon-Spin Contribution to the Proton Spin from the Double-Helicity Asymmetry in Inclusive $\pi^0$ Production in Polarized $p + p$ Collisions at $\sqrt{s}=200$ GeV,
\href{https://doi.org/10.1103/PhysRevLett.103.012003}{Phys. Rev. Lett. \textbf{103}, 012003 (2009)}.

\bibitem{STAR:2013zyt}
L.~Adamczyk \textit{et al.} (STAR Collaboration),
Neutral pion cross section and spin asymmetries at intermediate pseudorapidity in polarized proton collisions at $\sqrt{s} = 200$ GeV,
\href{https://doi.org/10.1103/PhysRevD.89.012001}{Phys. Rev. D \textbf{89}, no.1, 012001 (2014)}.

\bibitem{PHENIX:2014gbf}
A.~Adare \textit{et al.} (PHENIX Collaboration),
Inclusive double-helicity asymmetries in neutral-pion and eta-meson production in $\vec{p}+\vec{p}$ collisions at $\sqrt{s}=200$ GeV,
\href{https://doi.org/10.1103/PhysRevD.90.012007}{Phys. Rev. D \textbf{90}, no.1, 012007 (2014)}.

\bibitem{PHENIX:2010aru}
A.~Adare \textit{et al.} (PHENIX Collaboration),
Event structure and double helicity asymmetry in jet production from polarized $p+p$ collisions at $\sqrt{s} = 200$ GeV,
\href{https://doi.org/10.1103/PhysRevD.84.012006}{Phys. Rev. D \textbf{84}, 012006 (2011)}.

\bibitem{STAR:2012hth}
L.~Adamczyk \textit{et al.} (STAR Collaboration),
Longitudinal and transverse spin asymmetries for inclusive jet production at mid-rapidity in polarized $p+p$ collisions at $\sqrt{s}=200$ GeV,
\href{https://doi.org/10.1103/PhysRevD.86.032006}{Phys. Rev. D \textbf{86}, 032006 (2012)}.

\bibitem{STAR:2014wox}
L.~Adamczyk \textit{et al.} (STAR Collaboration),
Precision Measurement of the Longitudinal Double-spin Asymmetry for Inclusive Jet Production in Polarized Proton Collisions at $\sqrt{s}=200$ GeV,
\href{https://doi.org/10.1103/PhysRevLett.115.092002}{Phys. Rev. Lett. \textbf{115}, no.9, 092002 (2015)}.

\bibitem{Collins:1989gx}
J.~C.~Collins, D.~E.~Soper, and G.~F.~Sterman,
Factorization of Hard Processes in QCD,
\href{https://doi.org/10.1142/9789814503266\_0001}{Adv. Ser. Direct. High Energy Phys. \textbf{5}, 1-91 (1989)}.

\bibitem{Collins:2011zzd}
J.~Collins,
\href{https://doi.org/10.1017/9781009401845}{\it Foundations of Perturbative QCD},
Cambridge Monographs on Particle Physics, Nuclear Physics and Cosmology Series Vol. 32 (Cambridge University Press, Cambridge, England, 2011).

\bibitem{deFlorian:2008mr}
D.~de Florian, R.~Sassot, M.~Stratmann, and W.~Vogelsang,
Global Analysis of Helicity Parton Densities and Their Uncertainties,
\href{https://doi.org/10.1103/PhysRevLett.101.072001}{Phys. Rev. Lett. \textbf{101}, 072001 (2008)}.

\bibitem{deFlorian:2009vb}
D.~de Florian, R.~Sassot, M.~Stratmann, and W.~Vogelsang,
Extraction of spin-dependent parton densities and their uncertainties,
\href{https://doi.org/10.1103/PhysRevD.80.034030}{Phys. Rev. D \textbf{80}, 034030 (2009)}.

\bibitem{deFlorian:2014yva}
D.~de Florian, R.~Sassot, M.~Stratmann, and W.~Vogelsang,
Evidence for Polarization of Gluons in the Proton,
\href{https://doi.org/10.1103/PhysRevLett.113.012001}{Phys. Rev. Lett. \textbf{113}, no.1, 012001 (2014)}.

\bibitem{Nocera:2014gqa}
E.~R.~Nocera \textit{et al.} (NNPDF Collaboration),
A first unbiased global determination of polarized PDFs and their uncertainties,
\href{https://doi.org/10.1016/j.nuclphysb.2014.08.008}{Nucl. Phys. B \textbf{887}, 276 (2014)}.

\bibitem{Blumlein:2002qeu}
J.~Blumlein and H.~Bottcher,
QCD analysis of polarized deep inelastic data and parton distributions,
\href{https://doi.org/10.1016/S0550-3213(02)00342-5}{Nucl. Phys. B \textbf{636}, 225 (2002)}.

\bibitem{Blumlein:2010rn}
J.~Blumlein and H.~Bottcher,
QCD Analysis of polarized deep inelastic scattering data,
\href{https://doi.org/10.1016/j.nuclphysb.2010.08.005}{Nucl. Phys. B \textbf{841}, 205 (2010)}.

\bibitem{Hirai:2008aj}
M.~Hirai \textit{et al.} (Asymmetry Analysis Collaboration),
Determination of gluon polarization from deep inelastic scattering and collider data,
\href{https://doi.org/10.1016/j.nuclphysb.2008.12.026}{Nucl. Phys. B \textbf{813}, 106 (2009)}.

\bibitem{Leader:2010rb}
E.~Leader, A.~V.~Sidorov, and D.~B.~Stamenov,
Determination of polarized parton densities from a QCD analysis of inclusive and semi-inclusive deep inelastic scattering data,
\href{https://doi.org/10.1103/PhysRevD.82.114018}{Phys. Rev. D \textbf{82}, 114018 (2010)
}.

\bibitem{Sato:2016tuz}
N.~Sato \textit{et al.} (Jefferson Lab Angular Momentum),
Iterative Monte Carlo analysis of spin-dependent parton distributions,
\href{https://doi.org/10.1103/PhysRevD.93.074005}{Phys. Rev. D \textbf{93}, no.7, 074005 (2016)}.

\bibitem{Ethier:2017zbq}
J.~J.~Ethier, N.~Sato, and W.~Melnitchouk,
First simultaneous extraction of spin-dependent parton distributions and fragmentation functions from a global QCD analysis,
\href{https://doi.org/10.1103/PhysRevLett.119.132001}{Phys. Rev. Lett. \textbf{119}, no.13, 132001 (2017)}.

\bibitem{Brodsky:1994kg}
S.~J.~Brodsky, M.~Burkardt, and I.~Schmidt,
Perturbative QCD constraints on the shape of polarized quark and gluon distributions,
\href{https://doi.org/10.1016/0550-3213(95)00009-H}{Nucl. Phys. B \textbf{441}, 197 (1995)}.

\bibitem{Avakian:2007xa}
H.~Avakian, S.~J.~Brodsky, A.~Deur, and F.~Yuan,
Effect of Orbital Angular Momentum on Valence-Quark Helicity Distributions,
\href{https://doi.org/10.1103/PhysRevLett.99.082001}{Phys. Rev. Lett. \textbf{99}, 082001 (2007)}.

\bibitem{Leader:2001kh}
E.~Leader, A.~V.~Sidorov, and D.~B.~Stamenov,
A New evaluation of polarized parton densities in the nucleon,
\href{https://doi.org/10.1007/s100520200901}{Eur. Phys. J. C \textbf{23}, 479 (2002)}.

\bibitem{Bourrely:2001du}
C.~Bourrely, J.~Soffer, and F.~Buccella,
A Statistical approach for polarized parton distributions,
\href{https://doi.org/10.1007/s100520100855}{Eur. Phys. J. C \textbf{23}, 487 (2002)}.

\bibitem{Liu:2019vsn}
T.~Liu, R.~S.~Sufian, G.~F.~de T\'eramond, H.~G.~Dosch, S.~J.~Brodsky, and A.~Deur,
Unified Description of Polarized and Unpolarized Quark Distributions in the Proton,
\href{https://doi.org/10.1103/PhysRevLett.124.082003}{Phys. Rev. Lett. \textbf{124}, no.8, 082003 (2020)}.

\bibitem{Jimenez-Delgado:2014xza}
P.~Jimenez-Delgado \textit{et al.} (Jefferson Lab Angular Momentum Collaboration),
Constraints on spin-dependent parton distributions at large $x$ from global QCD analysis,
\href{https://doi.org/10.1016/j.physletb.2014.09.049}{Phys. Lett. B \textbf{738}, 263 (2014)}.

\bibitem{Roberts:2013mja}
C.~D.~Roberts, R.~J.~Holt, and S.~M.~Schmidt,
Nucleon spin structure at very high $x$,
\href{https://doi.org/10.1016/j.physletb.2013.09.038}{Phys. Lett. B \textbf{727}, 249 (2013)}.

\bibitem{Yu:2024qsd}
Y.~Yu, P.~Cheng, H.~Y.~Xing, F.~Gao, and C.~D.~Roberts,
Contact interaction study of proton parton distributions,
\href{https://doi.org/10.1140/epjc/s10052-024-13068-y}{Eur. Phys. J. C \textbf{84}, no.7, 739 (2024)}.

\bibitem{Ma:1996np}
B.-Q.~Ma,
The $x$-dependent helicity distributions for valence quarks in nucleons,
\href{https://doi.org/10.1016/0370-2693(96)00208-0}{Phys. Lett. B \textbf{375}, 320 (1996)}
[Erratum: \href{https://doi.org/10.1016/0370-2693(96)00631-4}{Phys. Lett. B \textbf{380}, 494 (1996)}].

\bibitem{Ma:1997gy}
B.-Q.~Ma, I.~Schmidt, and J.~Soffer,
The Quark spin distributions of the nucleon,
\href{https://doi.org/10.1016/S0370-2693(98)01158-7}{Phys. Lett. B \textbf{441}, 461 (1998)}.

\bibitem{Pasquini:2008ax}
B.~Pasquini, S.~Cazzaniga, and S.~Boffi,
Transverse momentum dependent parton distributions in a light-cone quark model,
\href{https://doi.org/10.1103/PhysRevD.78.034025}{Phys. Rev. D \textbf{78}, 034025 (2008)}.

\bibitem{Bacchetta:2008af}
A.~Bacchetta, F.~Conti, and M.~Radici,
Transverse-momentum distributions in a diquark spectator model,
\href{https://doi.org/10.1103/PhysRevD.78.074010}{Phys. Rev. D \textbf{78}, 074010 (2008)}.

\bibitem{Wigner:1939cj}
E.~P.~Wigner,
On Unitary Representations of the Inhomogeneous Lorentz Group,
\href{https://doi.org/10.2307/1968551}{Annals Math. \textbf{40}, 149 (1939)}.

\bibitem{Melosh:1974cu}
H.~J.~Melosh,
Quarks: Currents and constituents,
\href{https://doi.org/10.1103/PhysRevD.9.1095}{Phys. Rev. D \textbf{9}, 1095 (1974)}.

\bibitem{Ma:1991xq}
B.-Q.~Ma,
Melosh rotation: Source for the proton's missing spin,
\href{https://doi.org/10.1088/0954-3899/17/5/001}{J. Phys. G \textbf{17}, L53-L58 (1991)}.

\bibitem{Ma:1992sj}
B.-Q.~Ma and Q.-R.~Zhang,
The Proton spin and the Wigner rotation,
\href{https://doi.org/10.1007/BF01557707}{Z. Phys. C \textbf{58}, 479 (1993)}.

\bibitem{Feynman:1969ej}
R.~P.~Feynman,
Very high-energy collisions of hadrons,
\href{https://doi.org/10.1103/PhysRevLett.23.1415}{Phys. Rev. Lett. \textbf{23}, 1415 (1969)}.

\bibitem{Ma:1996ii}
B.-Q.~Ma and A.~Schafer,
Parton sum rules and improved scaling variable,
\href{https://doi.org/10.1016/0370-2693(96)00392-9}{Phys. Lett. B \textbf{378}, 307 (1996)}
[Erratum: \href{https://doi.org/10.1016/0370-2693(96)00630-2}{Phys. Lett. B \textbf{380}, 495 (1996)}]

\bibitem{Bacchetta:2017gcc}
A.~Bacchetta, F.~Delcarro, C.~Pisano, M.~Radici, and A.~Signori,
Extraction of partonic transverse momentum distributions from semi-inclusive deep-inelastic scattering, Drell-Yan and Z-boson production,
\href{https://doi.org/10.1007/JHEP06(2017)081}{J. High Energy Phys. 06 (2017) 081}
[Erratum: \href{https://doi.org/10.1007/JHEP06(2019)051}{J. High Energy Phys. 06 (2019) 051}].

\bibitem{Sun:2014dqm}
P.~Sun, J.~Isaacson, C.~P.~Yuan, and F.~Yuan,
Nonperturbative functions for SIDIS and Drell\textendash{}Yan processes,
\href{https://doi.org/10.1142/S0217751X18410063}{Int. J. Mod. Phys. A \textbf{33}, no.11, 1841006 (2018)}.

\bibitem{Scimemi:2019cmh}
I.~Scimemi and A.~Vladimirov,
Non-perturbative structure of semi-inclusive deep-inelastic and Drell-Yan scattering at small transverse momentum,
\href{https://doi.org/10.1007/JHEP06(2020)137}{J. High Energy Phys. 06 (2020) 137}.

\bibitem{Moos:2023yfa}
V.~Moos, I.~Scimemi, A.~Vladimirov, and P.~Zurita,
Extraction of unpolarized transverse momentum distributions from the fit of Drell-Yan data at N$^{4}$LL,
\href{https://doi.org/10.1007/JHEP05(2024)036}{J. High Energy Phys. 05 (2024) 036}.

\bibitem{Kang:2015msa}
Z.~B.~Kang, A.~Prokudin, P.~Sun, and F.~Yuan,
Extraction of quark transversity distribution and Collins fragmentation functions with QCD evolution,
\href{https://doi.org/10.1103/PhysRevD.93.014009}{Phys. Rev. D \textbf{93}, no.1, 014009 (2016)}.

\bibitem{Bacchetta:2020gko}
A.~Bacchetta, F.~Delcarro, C.~Pisano, and M.~Radici,
The 3-dimensional distribution of quarks in momentum space,
\href{https://doi.org/10.1016/j.physletb.2022.136961}{Phys. Lett. B \textbf{827}, 136961 (2022)}.

\bibitem{Gamberg:2022kdb}
L.~Gamberg \textit{et al.} (Jefferson Lab Angular Momentum Collaboration),
Updated QCD global analysis of single transverse-spin asymmetries: Extracting $\tilde{H}$, and the role of the Soffer bound and lattice QCD,
\href{https://doi.org/10.1103/PhysRevD.106.034014}{Phys. Rev. D \textbf{106}, no.3, 034014 (2022)}.

\bibitem{Bury:2021sue}
M.~Bury, A.~Prokudin, and A.~Vladimirov,
Extraction of the Sivers function from SIDIS, Drell-Yan, and $W^\pm/Z$ boson production data with TMD evolution,
\href{https://doi.org/10.1007/JHEP05(2021)151}{J. High Energy Phys. 05 (2021) 151}.

\bibitem{Echevarria:2020hpy}
M.~G.~Echevarria, Z.-B.~Kang, and J.~Terry,
Global analysis of the Sivers functions at NLO+NNLL in QCD,
\href{https://doi.org/10.1007/JHEP01(2021)126}{J. High Energy Phys. 01 (2021) 126}.

\bibitem{Zeng:2022lbo}
C.~Zeng, T.~Liu, P.~Sun, and Y.~Zhao,
Toward three-dimensional nucleon structures at the Electron-Ion Collider in China: A study of the Sivers function,
\href{https://doi.org/10.1103/PhysRevD.106.094039}{Phys. Rev. D \textbf{106}, no.9, 094039 (2022)}.

\bibitem{Boglione:2024dal}
M.~Boglione, U.~D'Alesio, C.~Flore, J.~O.~Gonzalez-Hernandez, F.~Murgia, and A.~Prokudin,
Simultaneous reweighting of Transverse Momentum Dependent distributions,
\href{https://doi.org/10.1016/j.physletb.2024.138712}{Phys. Lett. B \textbf{854}, 138712 (2024)}.

\bibitem{Zeng:2023nnb}
C.~Zeng, H.~Dong, T.~Liu, P.~Sun, and Y.~Zhao,
Role of sea quarks in the nucleon transverse spin,
\href{https://doi.org/10.1103/PhysRevD.109.056002}{Phys. Rev. D \textbf{109}, no.5, 056002 (2024)}.

\bibitem{Christova:2020ahe}
E.~Christova, D.~Kotlorz, and E.~Leader,
New study of the Boer-Mulders function: Implications for the quark and hadron transverse momenta,
\href{https://doi.org/10.1103/PhysRevD.102.014035}{Phys. Rev. D \textbf{102}, no.1, 014035 (2020)}.

\bibitem{Bhattacharya:2021twu}
S.~Bhattacharya, Z.-B.~Kang, A.~Metz, G.~Penn, and D.~Pitonyak,
First global QCD analysis of the TMD $\textsl{g}_{1T}$ from semi-inclusive DIS data,
\href{https://doi.org/10.1103/PhysRevD.105.034007}{Phys. Rev. D \textbf{105}, no.3, 034007 (2022)}.

\bibitem{Horstmann:2022xkk}
M.~Horstmann, A.~Schafer, and A.~Vladimirov,
Study of the worm-gear-T function $\textsl{g}_{1T}$ with semi-inclusive DIS data,
\href{https://doi.org/10.1103/PhysRevD.107.034016}{Phys. Rev. D \textbf{107}, no.3, 034016 (2023)}.

\bibitem{Yang:2024bfz}
K.~Yang, T.~Liu, P.~Sun, Y.~Zhao, and B.-Q.~Ma,
Extraction of transhelicity worm-gear distributions and opportunities at the Electron-Ion Collider in China,
\href{https://doi.org/10.1103/PhysRevD.110.034036}{Phys. Rev. D \textbf{110}, no.3, 034036 (2024)}.

\bibitem{Angeles-Martinez:2015sea}
R.~Angeles-Martinez \textit{et al.},
Transverse Momentum Dependent (TMD) parton distribution functions: status and prospects,
\href{https://doi.org/10.5506/APhysPolB.46.2501}{Acta Phys. Polon. B \textbf{46}, no.12, 2501 (2015)}.

\bibitem{Boussarie:2023izj}
R.~Boussarie \textit{et al.},
TMD Handbook,
\href{https://doi.org/10.48550/arXiv.2304.03302}{\tt arXiv:2304.03302}.

\bibitem{HERMES:2018awh}
A.~Airapetian \textit{et al.} (HERMES Collaboration),
Longitudinal double-spin asymmetries in semi-inclusive deep-inelastic scattering of electrons and positrons by protons and deuterons,
\href{https://doi.org/10.1103/PhysRevD.99.112001}{Phys. Rev. D \textbf{99}, no.11, 112001 (2019)}.

\bibitem{COMPASS:2016klq}
C.~Adolph \textit{et al.} (COMPASS Collaboration),
Azimuthal asymmetries of charged hadrons produced in high-energy muon scattering off longitudinally polarised deuterons,
\href{https://doi.org/10.1140/epjc/s10052-018-6379-7}{Eur. Phys. J. C \textbf{78}, no.11, 952 (2018)}.


\bibitem{CLAS:2017yrm}
S.~Jawalkar \textit{et al.} (CLAS Collaboration),
Semi-Inclusive $\pi_0$ target and beam-target asymmetries from 6 GeV electron scattering with CLAS,
\href{https://doi.org/10.1016/j.physletb.2018.06.014}{Phys. Lett. B \textbf{782}, 662 (2018)}.

\bibitem{Dudek:2012vr}
J.~Dudek \textit{et al.},
Physics Opportunities with the 12 GeV Upgrade at Jefferson Lab,
\href{https://doi.org/10.1140/epja/i2012-12187-1}{Eur. Phys. J. A \textbf{48}, 187 (2012)}.

\bibitem{Bacchetta:2004jz}
A.~Bacchetta, U.~D'Alesio, M.~Diehl, and C.~A.~Miller,
Single-spin asymmetries: The Trento conventions,
\href{https://doi.org/10.1103/PhysRevD.70.117504}{Phys. Rev. D \textbf{70}, 117504 (2004)}.

\bibitem{Collins:1981uk}
J.~C.~Collins and D.~E.~Soper,
Back-To-Back Jets in QCD,
\href{https://doi.org/10.1016/0550-3213(81)90339-4}{Nucl. Phys. B \textbf{193}, 381 (1981)}
[Erratum: \href{https://doi.org/10.1016/0550-3213(83)90235-3}{Nucl. Phys. B \textbf{213}, 545 (1983)}].

\bibitem{Ji:2004wu}
X.~d.~Ji, J.~p.~Ma, and F.~Yuan,
QCD factorization for semi-inclusive deep-inelastic scattering at low transverse momentum,
\href{https://doi.org/10.1103/PhysRevD.71.034005}{Phys. Rev. D \textbf{71}, 034005 (2005)}.

\bibitem{Ji:2004xq}
X.~d.~Ji, J.~P.~Ma, and F.~Yuan,
QCD factorization for spin-dependent cross sections in DIS and Drell-Yan processes at low transverse momentum,
\href{https://doi.org/10.1016/j.physletb.2004.07.026}{Phys. Lett. B \textbf{597}, 299 (2004)}.

\bibitem{Aybat:2011zv}
S.~M.~Aybat and T.~C.~Rogers,
TMD Parton Distribution and Fragmentation Functions with QCD Evolution,
\href{https://doi.org/10.1103/PhysRevD.83.114042}{Phys. Rev. D \textbf{83}, 114042 (2011)}.

\bibitem{Scimemi:2018xaf}
I.~Scimemi and A.~Vladimirov,
Systematic analysis of double-scale evolution,
\href{https://doi.org/10.1007/JHEP08(2018)003}{J. High Energy Phys. 08 (2018) 003}.

\bibitem{Collins:1981va}
J.~C.~Collins and D.~E.~Soper,
Back-To-Back Jets: Fourier Transform from B to K-Transverse,
\href{https://doi.org/10.1016/0550-3213(82)90453-9}{Nucl. Phys. B \textbf{197}, 446 (1982)}.

\bibitem{Gutierrez-Reyes:2017glx}
D.~Guti\'errez-Reyes, I.~Scimemi, and A.~A.~Vladimirov,
Twist-2 matching of transverse momentum dependent distributions,
\href{https://doi.org/10.1016/j.physletb.2017.03.031}{Phys. Lett. B \textbf{769}, 84 (2017)}.

\bibitem{NNPDF:2017mvq}
R.~D.~Ball \textit{et al.} (NNPDF Collaboration),
Parton distributions from high-precision collider data,
\href{https://doi.org/10.1140/epjc/s10052-017-5199-5}{Eur. Phys. J. C \textbf{77}, no.10, 663 (2017)}.

\bibitem{deFlorian:2014xna}
D.~de~Florian, R.~Sassot, M.~Epele, R.~J.~Hern\'andez-Pinto,  and M.~Stratmann,
Parton-to-Pion Fragmentation Reloaded,
\href{https://doi.org/10.1103/PhysRevD.91.014035}{Phys. Rev. D \textbf{91}, 014035 (2015)}
[\href{https://doi.org/10.48550/arXiv.1410.6027}{\tt arXiv:1410.6027}].

\bibitem{deFlorian:2017lwf}
D.~de~Florian, M.~Epele, R.~J.~Hern\'andez-Pinto, R.~Sassot and M.~Stratmann,
Parton-to-Kaon Fragmentation Revisited,
\href{https://doi.org/10.1103/PhysRevD.95.094019}{Phys. Rev. D \textbf{95}, 094019 (2017)}
[\href{https://doi.org/10.48550/arXiv.1702.06353}{\tt arXiv:1702.06353}].

\bibitem{deFlorian:2007ekg}
D.~de~Florian, R.~Sassot and M.~Stratmann,
Global analysis of fragmentation functions for protons and charged hadrons,
\href{https://doi.org/10.1103/PhysRevD.76.074033}{Phys. Rev. D \textbf{76}, 074033 (2007)}
[\href{https://doi.org/10.48550/arXiv.0707.1506}{\tt arXiv:0707.1506}].

\bibitem{sup}
See Supplemental Material at \href{http://link.aps.org/supplemental/10.1103/PhysRevLett.134.121902}{http://link.aps.org/\newline supplemental/10.1103/PhysRevLett.134.121902} for the
results of parameters, comparison with experimental data, the positivity bound test, and the fit with collinear distribution as input.

\bibitem{Echevarria:2011epo}
M.~G.~Echevarria, A.~Idilbi, and I.~Scimemi,
Factorization Theorem For Drell-Yan At Low $q_T$ And Transverse Momentum Distributions On-The-Light-Cone,
\href{https://doi.org/10.1007/JHEP07(2012)002}{J. High Energy Phys. 07 (2012) 002}.

\bibitem{Ebert:2022cku}
M.~A.~Ebert, J.~K.~L.~Michel, I.~W.~Stewart, and Z.~Sun,
Disentangling long and short distances in momentum-space TMDs,
\href{https://doi.org/10.1007/JHEP07(2022)129}{J. High Energy Phys. 07 (2022) 129}.

\bibitem{Bacchetta:2024yzl}
A.~Bacchetta, A.~Bongallino, M.~Cerutti, M.~Radici, and L.~Rossi, companion Letter, Exploring the three-dimensional momentum distribution of longitudinally polarized quarks in the proton, \href{https://journals.aps.org/prl/abstract/10.1103/PhysRevLett.134.121901}{Phys. Rev. Lett. 134, 121901 (2025)}.

\end{thebibliography}
\end{document}